# Imaging thick objects with deep-sub-angstrom resolution and deep-sub-picometer precision

Wenfeng Yang[1,2,3], Haozhi Sha[1,2,3], Jizhe Cui[1,2,3], Rong Yu[1,2,3]*

[1]School of Materials Science and Engineering, Tsinghua University, Beijing 100084, China.
[2]MOE Key Laboratory of Advanced Materials, Tsinghua University, Beijing 100084, China.
[3]State Key Laboratory of New Ceramics and Fine Processing, Tsinghua University, Beijing 100084, China.
*Corresponding author. Email: **ryu@tsinghua.edu.cn**

**Abstract:**

Size effects are ubiquitous in the structural, mechanical, and physical properties of materials, making it highly desirable to study the intrinsic properties of thick objects through high-resolution structural analysis in transmission electron microscopy. Although deep-sub-angstrom resolution has been achieved with multislice electron ptychography, the sample thickness is typically very limited. By combining energy filtering and extended local-orbital ptychography (eLOP) that retrieves varying aberrations during electron scanning, here we report ptychographic reconstructions for silicon as thick as 85 nm, approximately three times larger than usual thickness threshold for conventional multislice electron ptychography. The elimination of aberration variations contributes to accurate reconstructions with an information limit of 18 pm and atomic position precision of 0.39 pm. Accurate ptychographic reconstructions for thick objects can facilitate the discovery or interpretation of intrinsic structural and physical phenomena in solids, which is of great significance in physics, chemistry, materials science, and semiconductor device engineering.

Transmission electron microscopy [1] is a widely used technique for real-space imaging of solids at high spatial resolution. With the development of ptychographic algorithms [2-7], transmission electron microscopy is increasingly used with ptychographic phase retrieval for ultrahigh spatial resolution and high phase accuracy. On a modern transmission electron microscope equipped with an aberration corrector [8,9], the information limit of electron ptychography reached 39 pm in 2018 [10], 23 pm in 2021 [11], and 14 pm in 2023 [7]. Recently, an information limit of 41 pm for 2D materials was achieved using an uncorrected electron microscope [12]. Deep-subangstrom-resolution (< 0.5 Å) imaging with high phase accuracy helps to accurately characterize the microstructure of materials [13], e.g., orientation mapping around dislocation kinks in a metal oxide [14], lattice-resolved antiferromagnetic imaging [15], subangstrom resolution imaging of beam-sensitive materials [16-18], measuring atomic positions with sub-picometer precision in a polar metal [19], measuring the concentration of oxygen vacancies in a superconductor [20], imaging polar vortex in twist $SrTiO_3$ [21], and counting and positioning single atoms of interstitial oxygen in a metal [22].

Atomic resolution imaging usually requires very thin samples, typically less than 10 nm for aberration-corrected TEM imaging and 20 nm for STEM imaging. High-quality ptychographic reconstructions also necessitate a sample thickness less than 20 nm for dense solids such as metals, ceramics, and semiconductors. However, size effects are pervasive in the structural, mechanical, and physical properties of materials. The reduction of sample thickness often results in altered structural or dynamical behaviors due to strain relaxation, surface diffusion and damage, or even phase transitions. Excessively thin samples pose significant challenges for the reliability of structural analysis in transmission electron microscopy. In order to obtain intrinsic structures of solids, it is highly desirable to use samples that are as thick as possible for accurate structural analysis.

In this study, we developed a method named extended local-orbital ptychography (eLOP) that enables the variation of aberrations as the electron beam scans across the sample. It should be noted that eLOP and its precursor (LOP) [7] are phase retrieval methods of unknown structures that starts from scratch. They differ from the method proposed in the reference [23], which involve the structural refinement of known initial models. By combining the eLOP method with energy-filtering, ptychographic reconstructions are realized for Si of 85 nm thick and SrTiO$_3$ of 60 nm thick, with the information limit of 18 pm for Si and 16 pm for SrTiO$_3$. The atomic position precisions for Si and SrTiO$_3$ are 0.39 pm and 0.42 pm, respectively, both reaching the deep-sub-picometer (< 0.5 pm) region.

Probe aberrations negatively affect high-resolution imaging of objects. In general, the problem is more severe for thick objects due to the larger focus range of the probe within the objects.

Conventional pixelated ptychography (CPP) describes the probe as pixel arrays, and optimizes the probe to eliminate the influence of aberrations, which are assumed to be constant during probe scanning. However, aberrations often vary in experiments. eLOP describes the probe as position-dependent aberration functions. See Methods for details.

Figure 1 shows the simulation tests conducted to evaluate the performance of the eLOP method in handling aberration variations, which are difficult to be considered in CPP [24] but are essential for ptychographic reconstructions with a varying probe. Fig. 1(a) and Fig. 1(b) show a probe with the aberrations dependent on the scanning positions. The variations in the real part of the aberration coefficients $A_1$ and $A_2$ used in the simulations are shown in the Fig. 1(b) and Fig. 1(c), respectively.

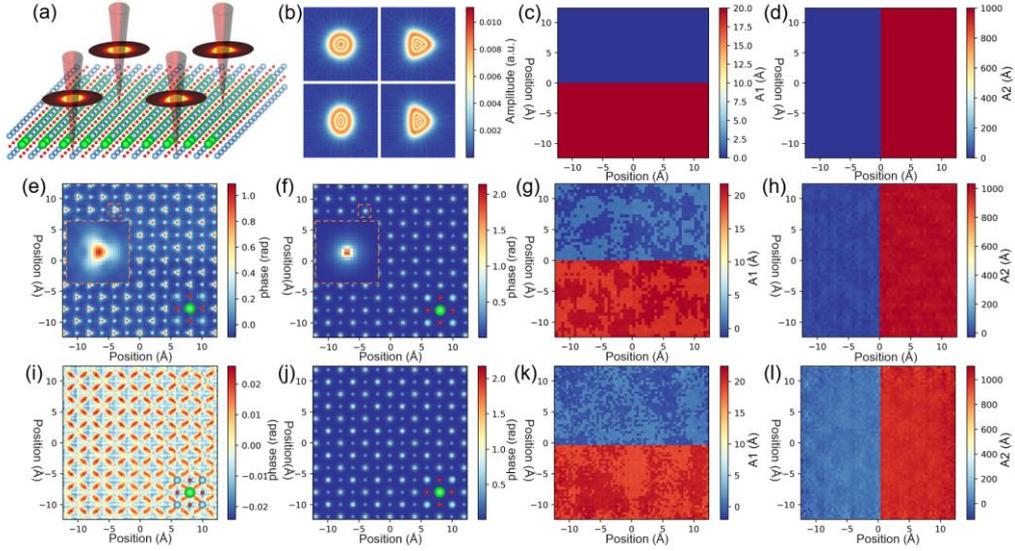

FIG. 1. Ptychographic reconstructions on simulated datasets with aberration variations. (a), schematic of the scanning probe with varying aberrations; (b), probe amplitude used in simulations corresponding to the aberrations shown in (c) and (d); (c), the real part of the aberration $A_1$ used in the simulations. (d), the real part of the aberration $A_2$ used in the simulations; (d), probe amplitude used in simulation corresponding to the aberrations shown in (b) and (c); (e), the object phase reconstructed via CPP for a 20 nm thick object. The atom in the inset is magnified five times; (f), the object phase reconstructed via eLOP for the object of 20 nm thick and the corresponding mapping of the aberration $A_1$ (g) and $A_2$ (h); i, the object phase reconstructed via CPP for the object of 60 nm thick; j, the object phase reconstructed via eLOP for a 60 nm thick object and the corresponding mapping of the aberration $A_1$ (k) and $A_2$ (l); The atomic structure is overlaid on the phase images. Sr, Ti and O atoms are shown with green, blue and red circles, respectively.

For a 20 nm thick object, the object phase reconstructed via CPP is shown in Fig. 1(d). The atom in the inset is magnified five times to clearly illustrate the irregular atomic shape clearly. Since the probe is independent of the scanning position, it can only account for the average aberrations in the scanning area. The residual aberrations in the probe enter the object phase, resulting in the irregular atomic shape and lattice deformation as shown in Fig. S1(a). To quantitatively evaluate the precision of atomic positions, the distance ($d_{Sr-TiO}$) between the TiO column and Sr column was measured. Fig. S2(a) shows the histogram of $d_{Sr-TiO}$ with a standard deviation of 2.65 pm. That means CPP is not able to remove the effect of aberration variations, which is reflected in the object phase. In contrast, as shown in Fig. 1(e), eLOP provides highly accurate shape and positions of atoms in the object phase as shown in Fig. S1(b). The standard deviation of $d_{Sr-TiO}$ is only 0.08 pm as shown in Fig. S2(b), due to the removal of aberration variations as shown in Fig. 1(f) and Fig. 1(g).

With increasing thickness, the influence of aberration variations on ptychographic reconstructions becomes stronger. For a 60 nm thick object, CPP fails to reconstruct the object phase, as shown in Fig. 1(h). In contrast, the object phase obtained by eLOP in Fig. 1(i) is accurate, with a standard deviation of $d_{Sr-TiO}$ measuring 0.34 pm, as shown in Fig. S2(c). The mapping of the aberrations in Fig. 1(j) and Fig. 1(k) are close to the true values used in the simulations.

Simulated datasets were used to demonstrate the ability of the eLOP method for the reconstruction of thick objects. For comparison, reconstructions using the CPP method were also performed. The multislice method [25] as implemented in the μSTEM code [26] was used. The 4D-STEM datasets for various thicknesses were generated, and ptychographic reconstructions were carried out. Typical diffraction patterns are shown in Fig. S3. Each diffraction pattern was padded to 256 × 256 pixels for better real-space sampling during the reconstructions. For the CPP method, we used the EMPTY program, which implements the multislice ptychography with adaptive propagator [27].

The maximum thickness reconstructed by eLOP is at least 60 nm, while by CPP is 40 nm, as shown in Fig. 2(a) and Fig. 2(b). The reason of larger thickness threshold of eLOP can be attributed to the robustness of the parameterized probe and object. The probe in CPP is composed of uncorrelated pixels. For objects of 50 nm and 60 nm thick, the probe cannot converge to a well-defined shape during the optimization. In contrast, the probe reconstructed via eLOP consistently maintains a well-defined shape, which benefits from the parameterization of the probe. Additionally, the object in eLOP is more robust than that in CPP. Although the object of 40 nm thick can be reconstructed via CPP, the sample thickness must be known and fixed during the reconstruction, otherwise it would diverge and results in unsuccessful reconstruction. In contrast, the sample thickness can be optimized with eLOP. This is a very desirable advantage since the sample thickness is usually unknown in experiments.

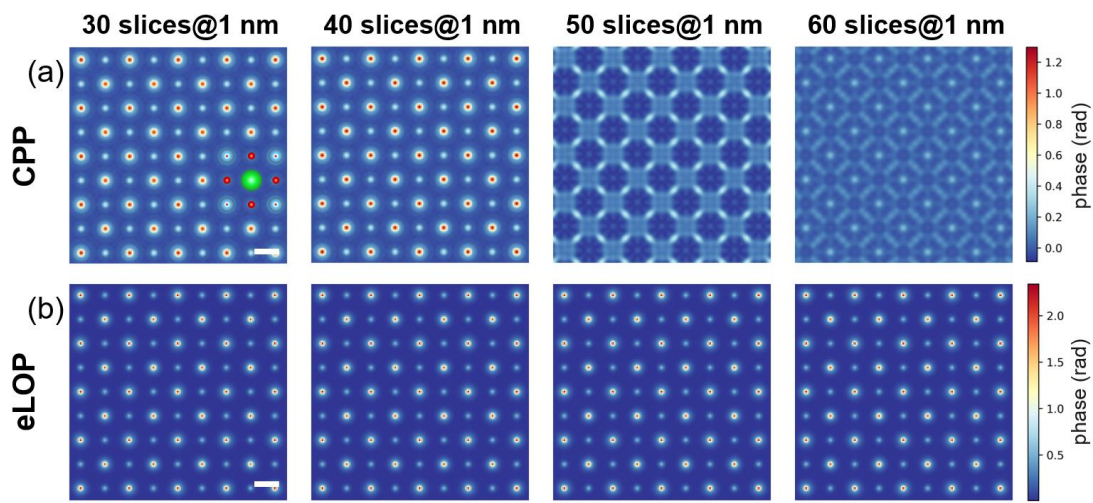

FIG. 2. Comparison of CPP and eLOP for different thicknesses on simulated datasets of $SrTiO_3$; (a) and (b), object phase and probe amplitude reconstructed via the CPP method; The object thicknesses are indicated at the top; The atomic structure is overlaid on the phase image; Sr, Ti and O atoms are shown with green, blue and red circles, respectively. Scale bar: (a), (b) 2 Å.

A $SrTiO_3$ sample with a thickness gradient was utilized to demonstrate the performance of the eLOP method to in enhancing the thickness threshold. Fig. 3(a) shows a low-magnification image of the sample. The thickness of the sample in different areas was estimated using low-loss Electron Energy Loss Spectroscopy (EELS), with the results shown in Fig. 3(b). Through the comparison between the true thickness derived from ptychographic reconstruction and the relative thickness (a value proportional to the inelastic mean free path, $\lambda$) estimated by EELS, the inelastic mean free path ($\lambda$) of $SrTiO_3$ was measured to be 115 nm. The elemental mapping is shown in Fig. S4. Energy-filtered 4D-STEM datasets used for ptychographic reconstructions were collected from the four areas indicated in the low-magnification image, as shown in Fig. 3(c). Typical diffraction patterns are shown in Fig. S5 and Fig. S6. Each diffraction pattern was padded to 256 × 256 pixels for fine real-space sampling during ptychographic reconstruction.

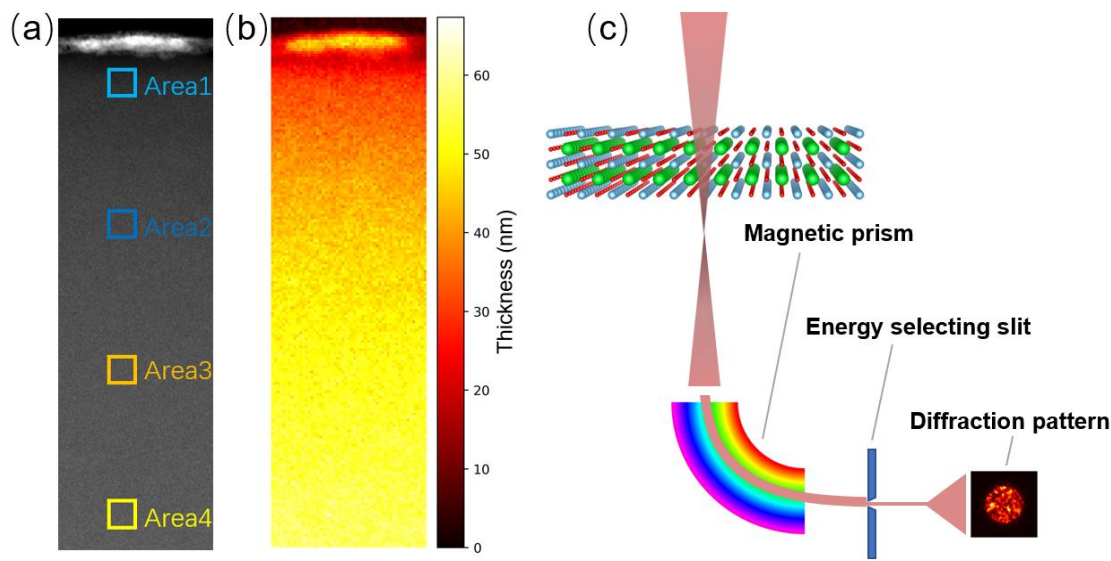

FIG. 3. Energy-filtering 4D-STEM experiment. (a), a low-magnification image of $SrTiO_3$ used to collect the experimental datasets. Four areas of different thicknesses for data collection are marked. (b), the thickness mapping by low-loss EELS, indicating areas 1-4 are approximately 30, 40, 50, and 60 nm thick, respectively. (c), Schematic of energy-filtering 4D-STEM experiment for ptychography.

Figure 4 shows the reconstructed phase images of different areas via different methods. The slice thickness is 1 nm in the reconstructions. For the 30 nm thick area, CPP shows good convergence for the energy-filtered dataset (EF-CPP) and fails to converge for the unfiltered dataset (UF-CPP), indicating the beneficial effect of removing inelastic scattering for ptychographic reconstructions. For thicker areas, CPP fails to converge for both filtered and unfiltered datasets. In contrast, eLOP shows good convergence for the areas up to 50 nm thick for the unfiltered datasets (UF-eLOP), and up to 60 nm for the filtered datasets (EF-eLOP). The improvement in thickness threshold of eLOP over CPP is remarkable.

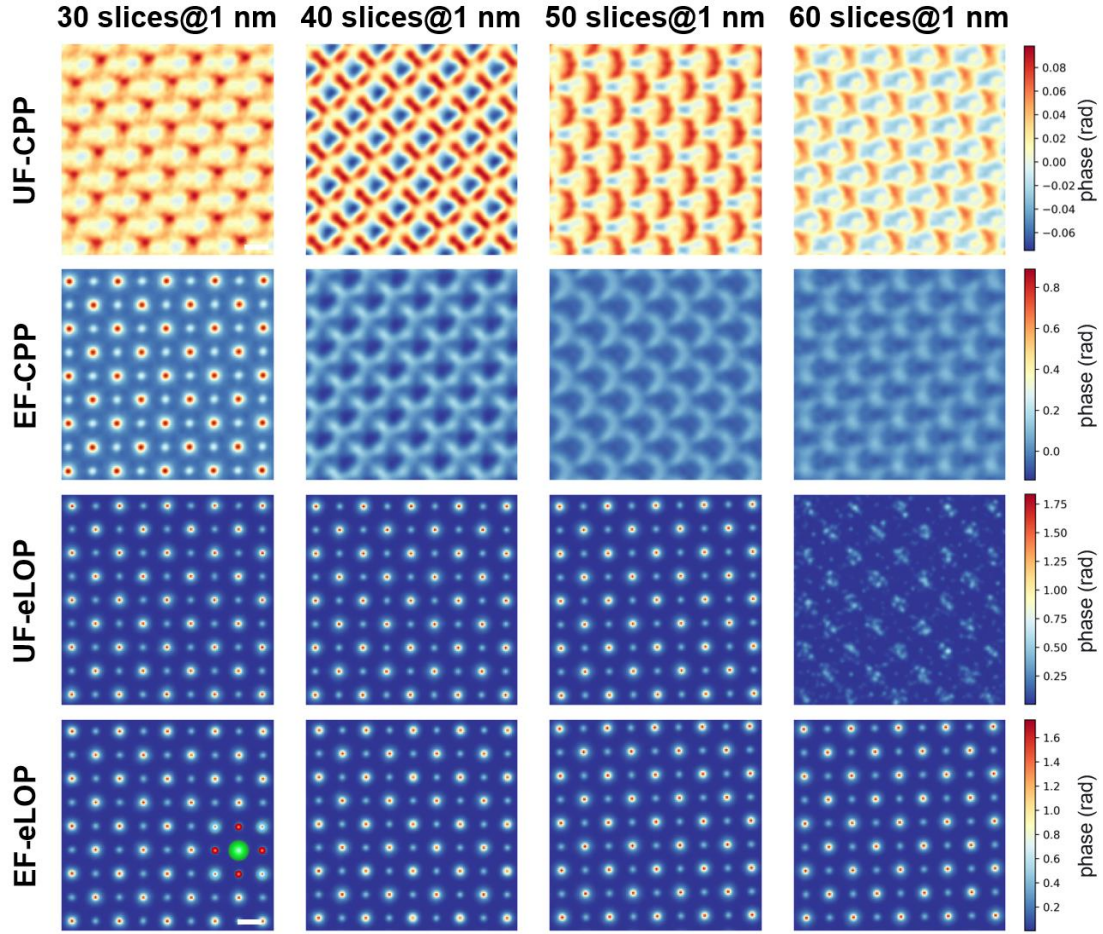

FIG. 4. Comparison of CPP and eLOP at different sample thickness for unfiltered and filtered diffraction patterns. The sample thicknesses are indicated at the top. The atomic structure is overlaid on the phase image. Sr, Ti and O atoms are shown with green, blue and red circles, respectively. Scale bar: 2 Å.

For the filtered dataset in the area of 60 nm thickness, eLOP also fails to converge when the position dependence of the probe is disabled, indicating the necessity of including the position-dependence of the probe aberrations for thick objects.

Figure 5 shows the information limit and the atom-position precision in ptychographic phase images of thick objects. The diffraction patterns are padded to 420 × 420 pixels for fine real-space sampling during reconstructions. The phase image averaged over the 60 slices of the 60-nm-thick $SrTiO_3$ is shown in Fig. 5(a). The information limit is 16 pm, as shown in Fig. 5(b). The distance $d_{Sr-TiO}$ between the Sr and TiO atomic columns is 272.2 pm, with a standard deviation of 0.42 pm, reaching the deep-sub-picometer region. The standard deviation of $d_{Sr-TiO}$ in the individual slices is about 0.6 to 1.1 pm, as shown in Fig. S7, indicating that thicker objects provide a high signal-to-noise ratio and is beneficial for high precision in atomic positions.

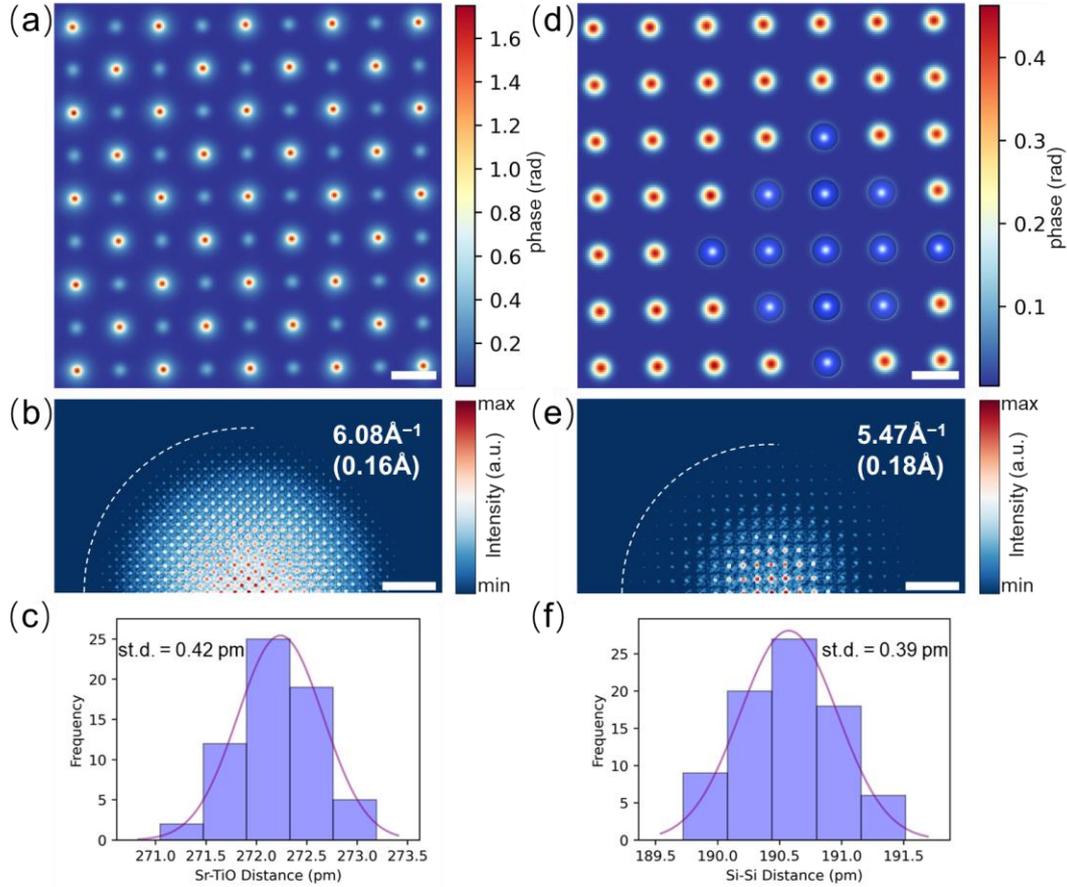

FIG. 5. The information limit and the atom-position precision in ptychographic phase images of thick objects. (a) the object phase of the 60-nm-thick $SrTiO_3$ along the [001] zone axis; (b) the diffractogram of (a); (c) the histogram of $d_{Sr-TiO}$ with a standard deviation of 0.42 pm; (d) the object phase of the 85-nm-thick Si along the [001] zone axis. (e) the diffractogram of (d); (f) the histogram of $d_{Si-Si}$ with a standard deviation of 0.39 pm. The atomic structure is overlaid on the phase image. Scale bar: (a), (d) 2 Å; (b), (e) 2 Å$^{-1}$.

Silicon, as the most important semiconductor, is used to further evaluate the thickness threshold of eLOP. Typical experimental CBED patterns and the PACBED are shown in Fig. S8. Figure 5(d) shows the reconstructed phase image of 85 nm thick Si along the [001] zone axis. The information limit is 18 pm, as shown in Fig. 5(e). The distance $d_{Si-Si}$ between the Si atomic columns is 190.6 pm, with a standard deviation of 0.39 pm, also reaching the deep-sub-picometer region.

By combining extended local-orbital ptychography and energy filtering, deep-sub-angstrom resolution and deep-sub-picometer precision have been achieved for dense solids as thick as 85 nm. This method alleviates the demanding requirements of electron ptychography on sample thickness, paving the way for ultrahigh-resolution imaging and accurate analysis of intrinsic structures and phenomena in engineering materials and semiconductor devices.

**Acknowledgements**

This work was supported by the National Natural Science Foundation of China (52388201). In this work we used the resources of the Physical Sciences Center and Center of High-Performance Computing of Tsinghua University.

**Author contributions Statement:**

R.Y. designed and supervised the research. W.Y. wrote the codes, performed experiments and simulations. W.Y. and R.Y. co-wrote the paper. All authors discussed the results and commented on the paper.

# References

[1]	D. B. Williams and C. B. Carter, Transmission Electron Microscopy 2009).
[2]	J. M. Rodenburg and H. M. L. Faulkner, A phase retrieval algorithm for shifting illumination, Applied Physics Letters **85**, 4795 (2004).
[3]	P. Thibault, M. Dierolf, A. Menzel, O. Bunk, C. David, and F. Pfeiffer, High-Resolution Scanning X-ray Diffraction Microscopy, Science **321**, 379 (2008).
[4]	A. M. Maiden, M. J. Humphry, and J. M. Rodenburg, Ptychographic transmission microscopy in three dimensions using a multi-slice approach, Journal of the Optical Society of America A **29**, 1606 (2012).
[5]	P. Thibault and A. Menzel, Reconstructing state mixtures from diffraction measurements, Nature **494**, 68 (2013).
[6]	H. Sha, J. Cui, and R. Yu, Deep sub-angstrom resolution imaging by electron ptychography with misorientation correction, Science Advances **8**, eabn2275 (2022).
[7]	W. Yang, H. Sha, J. Cui, L. Mao, and R. Yu, Local-orbital ptychography for ultrahigh-resolution imaging, Nature Nanotechnology **19**, 612 (2024).
[8]	M. Haider, H. Rose, S. Uhlemann, E. Schwan, B. Kabius, and K. Urban, A spherical-aberration-corrected 200kV transmission electron microscope, Ultramicroscopy **75**, 53 (1998).
[9]	O. L. Krivanek, N. Dellby, and A. R. Lupini, Towards sub-Å electron beams, Ultramicroscopy **78**, 1 (1999).
[10]	Y. Jiang *et al.*, Electron ptychography of 2D materials to deep sub-ångström resolution, Nature **559**, 343 (2018).
[11]	Z. Chen *et al.*, Electron ptychography achieves atomic-resolution limits set by lattice vibrations, Science **372**, 826 (2021).
[12]	K. X. Nguyen, Y. Jiang, C. H. Lee, P. Kharel, Y. Zhang, A. M. van der Zande, and P. Y. Huang, Achieving sub-0.5-angstrom-resolution ptychography in an uncorrected electron microscope, Science **383**, 865 (2024).
[13]	R. Yu, H. Sha, J. Cui, and W. Yang, Introduction to electron ptychography for materials scientists, Microstructures **4**, 2024056 (2024).
[14]	H. Sha, Y. Ma, G. Cao, J. Cui, W. Yang, Q. Li, and R. Yu, Sub-nanometer-scale mapping of crystal orientation and depth-dependent structure of dislocation cores in SrTiO3, Nature Communications **14**, 162 (2023).
[15]	J. Cui, H. Sha, W. Yang, and R. Yu, Antiferromagnetic imaging via ptychographic phase retrieval, Science Bulletin **69**, 466 (2024).
[16]	Z. Dong, E. Zhang, Y. Jiang, Q. Zhang, A. Mayoral, H. Jiang, and Y. Ma, Atomic-Level Imaging of Zeolite Local Structures Using Electron Ptychography, Journal of the American Chemical Society **145**, 6628 (2023).
[17]	H. Sha, J. Cui, J. Li, Y. Zhang, W. Yang, Y. Li, and R. Yu, Ptychographic measurements of varying size and shape along zeolite channels, Science Advances **9**, eadf1151 (2023).
[18]	H. Zhang *et al.*, Three-dimensional inhomogeneity of zeolite structure and composition revealed by electron ptychography, Science **380**, 633 (2023).
[19]	J. Zhang *et al.*, A correlated ferromagnetic polar metal by design, Nature Materials **23**, 912 (2024).
[20]	Z. Dong *et al.*, Visualization of oxygen vacancies and self-doped ligand holes in

La(3)Ni(2)O(7-δ), Nature **630**, 847 (2024).

[21] H. Sha, Y. Zhang, Y. Ma, W. Li, W. Yang, J. Cui, Q. Li, H. Huang, and R. Yu, Polar vortex hidden in twisted bilayers of paraelectric SrTiO3, Nature Communications **15**, 10915, 10915 (2024).

[22] J. Cui, H. Sha, L. Mao, K. Sun, W. Yang, and R. Yu, Imaging, counting, and positioning single interstitial atoms in solids, arxiv (2024).

[23] B. Diederichs, Z. Herdegen, A. Strauch, F. Filbir, and K. Müller-Caspary, Exact inversion of partially coherent dynamical electron scattering for picometric structure retrieval, Nature Communications **15**, 101 (2024).

[24] M. Odstrcil, P. Baksh, S. A. Boden, R. Card, J. E. Chad, J. G. Frey, and W. S. Brocklesby, Ptychographic coherent diffractive imaging with orthogonal probe relaxation, Optics Express **24**, 8360 (2016).

[25] J. M. Cowley and A. F. Moodie, The scattering of electrons by atoms and crystals. I. A new theoretical approach, Acta Crystallographica **10**, 609 (1957).

[26] L. J. Allen, A. J. D'Alfonso, and S. D. Findlay, Modelling the inelastic scattering of fast electrons, Ultramicroscopy **151**, 11 (2015).

[27] H. Sha, J. Cui, and R. Yu, Deep sub-angstrom resolution imaging by electron ptychography with misorientation correction, Science Advances **8** (2022).